\documentclass{llncs}
\usepackage[T1]{fontenc}
\usepackage{graphicx}
\usepackage{xspace}

\usepackage{gospel}
\usepackage{why3lang}

\usepackage{subcaption}
\usepackage{graphicx}
\usepackage{stmaryrd, amssymb}

\usepackage{color, colortbl}
\usepackage{cellspace}
\usepackage{cite}
\usepackage{amssymb}
\usepackage{amsmath}
\usepackage{mathpartir}
\usepackage{ upgreek }
\usepackage{url}
\usepackage{xparse}

\usepackage{hyperref}

\DeclareCaptionFormat{listing}{\hfill#3}
\begin{document}

\newenvironment{judge}[2][c]
{\begin{minipage}[#1]{#2}
\hrule height2pt
\centerline{\vrule width2pt height 5pt\hfill \vrule width2pt height 5pt}
\begin{minipage}{\dimexpr\textwidth-4pt-1em}}
{\end{minipage}
\centerline{\vrule width2pt height 5pt\hfill \vrule width2pt height 5pt}
\hrule height2pt
\end{minipage}}

\newcommand{\why}{{Why3}\xspace}
\newcommand{\whyml}{{WhyML}\xspace}
\newcommand{\ocaml}{{OCaml}\xspace}
\newcommand{\vocal}{{VOCaL}\xspace}
\newcommand{\gospellang}{{GOSPEL}\xspace}
\newcommand{\dafnylang}{{Dafny}\xspace}
\newcommand{\mixins}{{Mixins}\xspace}
\newcommand{\coq}{{Coq}\xspace}
\newcommand{\iris}{{Iris}\xspace}
\newcommand{\clang}{{C}\xspace}
\newcommand{\cfml}{{CFML}\xspace}
\newcommand{\cameleer}{{Cameleer}\xspace}
\newcommand{\cps}{{CPS}\xspace}
\newcommand{\sepimp}{\mathrel{-\mkern-6mu*}}
\newcommand{\client}{\texttt{Client}\xspace}
\newcommand{\server}{\texttt{Server}\xspace}
\newcommand{\koda}{{Koda-Ruskey}\xspace}

\newcommand{\nbnote}[3]{
  \fcolorbox{gray}{yellow}{\bfseries\sffamily\scriptsize#1}
  {\color{#2} \sffamily\small$\blacktriangleright$\textit{#3}$\blacktriangleleft$}
}

\newcommand{\mario}[1]{\nbnote{Mario}{red}{#1}}
\renewcommand{\mario}[1]{}
\newcommand{\tiago}[1]{\nbnote{Tiago}{blue}{#1}} %
\renewcommand{\tiago}[1]{}

\newcommand{\myparagraph}[1]{\vspace{-10.5pt}\paragraph{#1}}
\newcommand{\mysection}[1]{\vspace{-2pt}\section{#1}}
\newcommand{\mycaption}[1]{\vspace{-7.5pt}\caption{#1}}

\renewcommand{\check}{
  \raisebox{-.25\height}{\includegraphics[width=1em,height=1em]{check.png}}}

\newcommand{\cross}{
  \raisebox{-.25\height}{\includegraphics[width=1em,height=1em]{cross.png}}}

\newcommand{\xbar}{\bar{x}}
\newcommand{\qqquad}{\qquad\qquad\quad}
\newcommand{\reg}{\texttt{reg}}
\newcommand{\btau}{\beta\tau}
\newcommand{\abar}{\bar{a}}
\newcommand{\btaubar}{\overline{\btau}}
\newcommand{\alphabar}{\overline{\alpha}}
\newcommand{\taubar}{\overline{\tau}}
\newcommand{\alabel}[1]{\quad\text{#1}}

\newcommand{\tprotocol}{(TProtocol)\xspace}
\newcommand{\teffect}{(TEffect)\xspace}
\newcommand{\tperform}{(TPerform)\xspace}
\newcommand{\tfun}{(TFun)\xspace}
\newcommand{\tperformsclause}{(TPerformsClause)\xspace}
\newcommand{\tappdefun}{(TAppDefun)\xspace}
\newcommand{\tapp}{(TApp)\xspace}
\newcommand{\ttry}{(TTry)\xspace}

\def \ifempty#1{\def\temp{#1} \ifx\temp\empty }
\newcommand{\exbar}{\overline{E\xbar\Rightarrow e}}
\newcommand{\betaxbar}{\overline{\beta x}}
\newcommand{\sep}{|\;}
\newcommand{\svars}{S_{\xbar}}
\newcommand{\sovars}{S_{old\xbar}}
\newcommand{\stype}{S_\tau}
\newcommand{\effenv}{\Upsigma}
\newcommand{\funenv}{\Updelta}
\newcommand{\argenv}{\nu}
\newcommand{\refenv}{\mu}
\newcommand{\env}[1]{\pi_#1(\effenv(E))}
\newcommand{\fullenv}[1]{\effenv#1\circ\funenv#1\circ\argenv#1\circ\refenv#1}
\newcommand{\allvars}{S_{vars}}
\newcommand{\sym}[2]{\ifthenelse{\equal{s}{#1}}{#2}{#1}}
\newcommand{\mkenv}[4]{
  \ifthenelse{\equal{}{#1}}{}{\sym{#1}{\effenv}\circ}
  \ifthenelse{\equal{}{#2}}{}
  {\sym{#2}{\funenv}\ifthenelse{\equal{}{#3}}{\ifthenelse{\equal{}{#4}}{}{\circ}}{\circ}}
  \ifthenelse{\equal{}{#3}}{}{\sym{#3}{\argenv}\ifthenelse{\equal{}{#4}}{}{\circ}}
  \sym{#4}{\refenv}
}
\newcommand{\s}{\mathcal{S}}
\newcommand{\df}{\mathcal{D}}
\newcommand{\h}{\mathcal{H}}
\newcommand{\iamgoingtokillyou}{\mathcal{O}}

\NewDocumentCommand{\trans}
{O{\fullenv{}} m m O{}}
{#1\vdash\llbracket#2\rrbracket\leadsto#3\dashv#4}

\definecolor{thegray}{rgb}{0.849,0.849,0.849}

\pagestyle{plain}

\title{
  A Framework for the Automated Verification of Algebraic Effects and Handlers\\
  (extended version)\thanks{This work is partly supported by the HORIZON 2020
    Cameleer project (Marie Skłodowska-Curie grant agreement ID:897873) and NOVA
    LINCS (Ref. UIDB/04516/2020)}}
\titlerunning{Automated Verification of Algebraic Effects}
%
\author{Tiago Soares \and Mário Pereira}
\authorrunning{T. Soares et al.}
%
\institute{NOVA LINCS, Nova School of Science and Technology, Portugal}

\maketitle              
\begin{abstract}
  Algebraic effects and handlers are a powerful abstraction to build non-local
  control-flow mechanisms such as resumable exceptions, lightweight threads,
  co-routines, generators, and asynchronous I/O. All of such features have very
  evolved semantics, hence they pose very interesting challenges to deductive
  verification techniques. In fact, there are very few proposed techniques to
  deductively verify programs featuring these constructs, even fewer when it
  comes to automated proofs. In this paper, we present an extension to
  \cameleer, a deductive verification tool for \ocaml code, that allows one to
  reason about algebraic effects and handlers. The proposed embeds the behavior
  of effects and handlers using exceptions and employs defunctionalization to
  deal with continuations exposed by effect handlers.

\end{abstract}

\section{Introduction}
\label{sec:introduction}

The newest release of the \ocaml compiler saw the introduction of a variety of
new features that make parallel and concurrent code more efficient as well as
easier to write. These include, amongst other, algebraic effects and
handlers~\cite{eff-t}. In short, algebraic effects are similar to exceptions:
when an effect is performed, execution is halted and control is surrendered to
the nearest enclosing handler. The main difference between exception and effect
handlers is that the latter exposes a (delimited) continuation function that
allows us to go back to the point in which the effect was performed and resume
execution.

This addition to an industrial strength language means there is a large
incentive for verification tools to add support for algebraic effects and
handlers. Although some research has been done~\cite{sep-eff, plot} that
presents reasoning rules for this kind of program, none of them explore
automated proofs. Therefore, the research question we pose to ourselves is
\emph{what tools and methodologies must be developed in order to apply automated
  deductive proofs to \ocaml programs, featuring effects and handlers?}

To answer this question, we present a framework that can automatically verify
programs that employ algebraic effects. Such a framework is implemented as an
extension to \cameleer~\cite{pereiracav21}, a deductive verification tool for
\ocaml code whose specification is written in
\gospellang~\cite{ChargueraudFLP19} (Generic \ocaml SPEcification Language) and
then translated to \whyml, the specification and programming language of the
\why verification framework~\cite{why}. \cameleer is also the only automated
deductive verification platform (we are aware of) that targets \ocaml programs,
making it an ideal choice for this project.

To achieve this goal we must be able to represent the continuation function
exposed by the handler in our proofs, which we will do using a translation of
algebraic effects into \why that employs defunctionalization~\cite{defunc}, a
program transformation technique that turns higher-order programs into
first-order ones. To the best of our knowledge, the use of defunctionalization
in a proof scenario is novel.

The contributions of this work can be summarized as follows:
\begin{enumerate}
\item we extend \gospellang to accommodate the needed logical elements to
  specify the behavior of algebraic effects and handlers;
\item we develop an embedding of \ocaml handlers and effects in \whyml,
  employing defunctionalization;
\item we extend \cameleer to automatically translate \gospellang annotated
  \ocaml programs into this embedding, allowing us to conduct deductive
  verification inside the \why framework; and
\item we build a set of case studies verified using our proposed methodology.
\end{enumerate}

\paragraph{Notation.} \ocaml and \gospellang share many syntactic traits with
\whyml. Since we show examples using these three languages throughout the paper,
we put a label on the right-hand side of every listing to indicate the language
in which such piece of code is written.

\mario{Note to self: mencionar o facto de usarmos etiquetas nos listings.}

\section{Algebraic Effects in a Nutshell}
\label{sec:algebraic-effects}

Before presenting how we plan to specify and prove algebraic effects, let us
present a program that simulates a mutable reference, without actually using
mutable state. For such purpose, we introduce the following two distinct
algebraic effects:

%
\begin{ocamlenv}
  effect Get : int
  effect Set : int -> unit
\end{ocamlenv}
The role of \whyf{Get} and \whyf{Set} effects is to implement, respectively, the
usual dereferencing and assignment operations.

One can read the above definitions as follows: when the \whyf{Get} effect is
performed, the nearest enclosing handler calls the corresponding continuation by
passing as argument an \whyf{int} value, which stands for the current value of
the reference. \mario{Stores? Holds?}
Conversely, the \whyf{Set} effect sends an \whyf{int}, the new value of the
reference, to the handler. The handler will simply return \whyf{unit}. This
means any future \whyf{Get} effect shall retrieve this same value.

Using these effects, we define the following function that increments the value
of the reference by one:
\begin{ocamlenv}
  let inc () : unit =
    let current = perform Get in
    perform (Set (1 + current))
\end{ocamlenv}
To call this function, we must define a handler that enforces the intended
behavior. Abiding to our rule of not using mutable state, we create the
following anonymous functions to represent the program state:
\begin{ocamlenv}
  let create_env () : int -> int = match inc () with
    | r ->                fun _ -> r
    | effect Get k ->     fun x -> (continue k x) x
    | effect (Set n) k -> fun _ -> (continue k ()) n
\end{ocamlenv}
Here, \whyf{create_env} returns another function which receives the initial
value of the reference and handles each effect accordingly. In case no effect is
performed, the handler falls into the first branch and it builds a function that
always returns the result~\whyf{r}.

\mario{explicar
  primeiro o ramo sem efeitos}
When the \whyf{Get} effect is handled we return a function that calls the
continuation \why{k} with the argument \whyf{x}, that is, the current value of
the reference. Important to note that when we call the continuation with the
\whyf{continue} function, the handler remains installed, meaning it handles any
remaining effects. Such kind of handlers is know as \emph{deep handlers},
whereas \emph{shallow handlers} can only handle at most one
effect~\cite{shallow}. This also means that calling the continuation returns
another function, seeing as it is the return type of the handler.
After the call to \whyf{continue}, we resume the computation back to the point
where the effect is performed, inside the \whyf{inc} function. As such, the
\whyf{Set} effect is performed (and handled by the next branch of the same
handler). This means \whyf{continue} returns a function, which we apply
to~\whyf{x} to ensure the value of the reference remains unchanged.

When the \whyf{Set} effect is handled, it returns a function that calls the
continuation, which once again returns a new function. In turn, this function is
applied to \whyf{n}, the value sent through the effect.


The use of \whyf{Get} and \whyf{Set} effects resemble the use of the state
monad, for instance, in the Haskell
language\footnote{\url{https://wiki.haskell.org/State_Monad}}. Indeed, algebraic
effects have many similarities with monads, in the sense that both constructions
provide abstraction over effectful computations. The upside of using effects is
that performing effects is done in \emph{direct style}, much like calling a
function. Whereas with monads, one would need to encapsulate every operation
inside the monad type. Moreover, effects provide a clear \emph{abstraction
  barrier} between the action of performing an effect and its actual
implementation~\cite{action}.

\section{Specifying and Reasoning About Effects and Handlers}
\label{sec:translation-scheme}

Proving the correctness of an handler leads to some specification and reasoning
challenges, the greatest of which begin proving that every call to
\whyf{continue} is sound. For the example of the previous Section, this would
mean proving that for every \whyf{Get} effect the handler replies with the
correct value. To do this, we must define some logical contract that the handler
must respect when calling the continuation. Additionally, the function that
performs the effect, must also respect a set of requirements to be able to
perform an effect.

Throughout this paper, we refer to any function that calls \whyf{continue} as
the \emph{\server}, whereas any function calling \whyf{perform} is designated
the \emph{\client}. The rules that these two agents must ensure when performing
or responding to an effect is described by
\emph{protocols}~\cite{sep-eff}. Protocols consist of preconditions (the
conditions \client must respect when performing an effect), postconditions (the
conditions that \server must respect when calling the continuation) and a
\whyf{modifies} clause with all the variables \server is given write permissions
to.

\mario{dizer que nesta secção vamos apresentar protocolos em Cameleer}

\subsection{Client Specification}
\label{sec:client}

To present how we implemented protocols in \cameleer, we use a simple example:
we create a program where \client sends some value to \server. The \server then
takes that value and sets a global reference~\whyf{p} to that value and returns
\whyf{p}'s old value. To do this, we first define the reference as well as an
effect, \whyf{XCHG}, that receives an integer value and returns another one. We
then create a protocol with the rules we outlined, as follows:
\begin{gospel}
  let p : int ref = ref 0
  effect XCHG : int -> int

  (*@ protocol XCHG x :
      ensures !p = x && reply = old !p
      modifies p *)
\end{gospel}
Here, we define a \gospellang protocol which ensures that, when \server returns
control to \client, the reference holds~\whyf{x}, the value passed by \client,
and \server replies with \whyf{p}'s old value (hence, the use of the keyword
\whyf{reply}). Additionally, we also state that \server is allowed to modify the
contents of \whyf{p}. This protocol has no precondition since there is no
restriction over what values \client may send. Every protocol is named after an
effect, meaning it establishes the rules the two agents must obey when using
such an effect to communicate.

Let us now move on to specifying a \client that uses this effect. This amounts
to the following:
\begin{gospel}
  let xchg (n : int) : int =
    perform (XCHG n)
  (*@ ensures !p = n && result = old !p
      performs XCHG *)
\end{gospel}
The \whyf{performs} clause lists the effects a certain function might
perform. The postcondition of the \whyf{xchg} function is exactly the
postcondition of the protocol, since it simply calls \whyf{perform} with effect
\whyf{XCHG}.

\tiago{escrever uma explicação da spec}







\subsection{Server Specification}
\label{sec:server-specification}

Consider the following simple \ocaml handler for the \whyf{XCHG} effect:
\begin{gospel}
  let server () =
    try xchg (xchg 42) with
    | effect (XCHG n) k ->
        let old_p = !p in
        p := n;
        continue k old_p
\end{gospel}
It is easy, at least informally, to observe that this handler respects the
protocol defined previously. However, in the general case, this can only be
established if we specify a condition whether the function exits through an
\emph{exceptional} branch or it ends normally. We refer to this piece of
specification as the \emph{handler invariant}, inspired from the fact we are
dealing here with deep handlers, which provide a form of recursive treatment of
effects. For the XGHG \server, we provide the following specification:
\begin{gospel}
  try xchg (xchg 42) with ...
  (*@ try_ensures !p = old !p && result = 42
      returns int *)
\end{gospel}
The said invariant is provided via \whyf{try_ensures}, a new clause we have
introduced in \gospellang for the purpose of reasoning about handlers. For the
above example, whenever the execution reaches the exceptional branch, we know
the final instruction of the handler is a call to \whyf{continue}. Since the
handler remains installed (again, due to it being a deep-handler)\mario{podemos
  dizer isto?}, it keeps calling \whyf{continue} until no more effects are
performed. In practice, this means the two effects produced by the double call
to \whyf{xchg} are treated in this handler. After the second call to
\whyf{continue}, function \whyf{server} halts its execution normally, hence the
handler specification allows one to derive this function returns~42 and the
value stored in~\whyf{p} is unchanged (even if this reference is written
twice). Finally, the \whyf{returns} keyword specifies the return type of this
function... \mario{porque é que precisamos disto aqui? E se assumirmos que
  estamos a trabalhar na AST tipada do OCaml?}






\section{Translation of Effects and Handlers into WhyML}
\label{sec:transl-effects-into}

Having presented our protocol-based approach to specify effects and handlers, we
focus now on exploiting such specification for the purpose of deductive
verification. We present in this Section a translation scheme from
\gospellang-annotated \ocaml programs, with effects and handlers, into a \whyml
program. Hence, the verification conditions generated by \why must ensure the
original \ocaml program adheres to its specification. This mainly amounts to
prove that \client and \server respect the established protocols. To guide our
presentation, we re-use the \whyf{xchg} example from
Sec.~\ref{sec:translation-scheme}.

\subsection{Reasoning about \client}
\label{sec:reas-about-client}

Each request \client makes, via an effect, must respect the corresponding
protocol's precondition and that \client's postcondition holds for any reply
\server may send. 
In order to prove such properties, our translation pipeline introduces the
following function, which simulates performing an \whyf{XCHG} effect:
\mario{talvez seja melhor usar `result` em todo o lado em vez de `reply`}
\begin{why3}
val perform_XCHG (x : int) : int
  ensures { let reply = result in
            reply = old !p && !p = x }
  writes  { p }
\end{why3}
The specification of \whyf{perform_XCHG} is exactly that of the \whyf{XCHG}
protocol. Using this function, the \client is straightforwardly translated into
\whyml, as follows:
\begin{why3}
  let xchg (n : int) : int
    ensures { !p = n && result = old !p }
  = perform_XCHG n
\end{why3}
By using the \whyf{perform_XCHG} function, we can check if the request made
satisfies the protocol's precondition and if the function's postcondition holds
for any possible reply. The former is proved trivially: seeing as there is no
precondition, \client can always make a request. As for the latter, the
protocols imposes \server can only reply after setting the~\whyf{p} reference to
the value sent by \client and can only return the reference's old value. This is
exactly what we put in \whyf{perform_XCHG} postcondition, hence the body of
\whyf{xchg} also adheres to its specification.

\subsection{Reasoning about \server}
\label{sec:reas-about-serv}

Using our translation approach, proving a \client is no different than proving
functions that call other functions. Proving \server, on the other hand, poses
some interesting challenges:
\begin{itemize}
  \item How does one translate \ocaml's effects into \whyml?
  \item The \why \whyf{xchg} function does not have any kind of exceptional
    return, it merely calls a function that simulates the handler's
    behavior. \mario{É mesmo o handler aqui?}
    How do we embed this in \why?
  \item How does one represent continuations, as well as \ocaml's
    \whyf{continue} function, in \why?
  \item What kind of specification should such continuations present?
\end{itemize}
To tackle the first issue, we simply choose to represent effects as \whyml
\emph{exceptions}. More specifically, we create an exception that receives the
same arguments as the original effect. In the case of \whyf{XCHG}, it receives a
single \whyf{int} value, resulting in the following encoding:
\begin{why3}
  exception XCHG int
\end{why3}
Using such representation for effects, \ocaml effect handlers are translated
into \whyml using exception handlers. The handler from
Sec.~\ref{sec:server-specification} is translated as follows:
\begin{why3}
  try xchg (xchg 42) with
  | XCHG n ->
      let old_p = !p in
      p := n;
      continue k old_p
\end{why3}
To make this handler effective, the \whyf{xchg} function must be able to,
indeed, throw an exception. Our translation scheme adds the following
\whyf{raises} clause to the \whyf{perform_XCHG} function:
\begin{why3}
  raises { XCHG }
\end{why3}
In practice, we translate every \whyf{performs} clause from \gospellang
specification into a \whyml \whyf{raises} clause, meaning that \whyf{xchg} will
have this same clause.


\subsection{Representing Continuations}
\label{sec:repr-cont-whyml}

The attentive reader might notice that the continuation~\whyf{k} is absent from
the \whyml translation of handlers. Indeed, representing and dealing with
continuations in \why is a challenge, since the framework has very limited
support for higher-order computation.

Continuations are, in essence, functions that can only be called using the
special \whyf{continue} function. Moreover, they can only be called once (if the
same continuation is called multiple times, the \ocaml runtime would throw an
exception). We encode continuations in \whyml encoding via the following record
type:
\begin{why3}
  type continuation 'a 'b = abstract {
    mutable _valid : bool;
  }
\end{why3}
This declaration introduces a polymorphic type of continuations, with argument
of type~\whyf{'a} and result of type~\of{'b}. The Boolean field \whyf{_valid} is
used to check whether the continuation was \emph{continue}d, ensuring the
one-shot behavior. From a programming perspective, type \whyf{continuation}
behaves as an \emph{abstract} data type. One can only explore the value of
\whyf{_valid} within specification or ghost code.

To use continuations as a value equipped with some logical contract, we adopt a
technique~\cite{Regis-GianasP08,KanigFilliatre09wml} that allows us to represent
and manipulate the specification of continuations. This is done through the
following pair of predicates:
\begin{why3}
  type state

  predicate pre  (continuation 'a 'b) 'a state
  predicate post (continuation 'a 'b) 'a state state 'b
\end{why3}
These predicates are not given a particular definition, but are axiomatized
later, as needed.
The \whyf{pre} predicate holds if the continuation's precondition holds. This
predicate is applied to the continuation's argument and a representation of the
state in which the continuation begins execution. On the other hand, \whyf{post}
predicate holds if the continuation's postcondition holds. This predicate is
applied to the continuation's argument, the initial state, the state after the
continuation executed, and finally the value returned by the continuation. The
\whyf{state} type is instantiated with the piece of mutable state that a
particular program manipulates. For instance, for the \whyf{xchg} example,
\whyf{state} would be refined as record type with a single mutable
field~\whyf{_p}, which represents the value stored in reference~\whyf{p}.

Using this approach, we define the \whyf{continue} function in \why, as follows:
\begin{why3}
  val continue (k : continuation 'a 'b) (arg : 'a) : 'b
    requires { k._valid }
    requires { pre k arg {_p = !p} }
    ensures  { post k arg (old {_p = !p}) {_p = !p} result }
    writes   { k._valid, p }
\end{why3}
When this function is called, it will respect the contract of
continuation~\whyf{k}. Additionally, it requires the continuation to be valid
(\emph{i.e.}, not called yet). Once \whyf{continue} returns, the continuation is
invalidated, since we state in the \whyf{writes} clause that the \whyf{k._valid}
field is written, hence one can no longer suppose it is set to \whyf{true}.

This encoding of the \whyf{continuation} data type is in fact a form of
defunctionalization~\cite{DanvyN01}, where \whyf{continue} resembles the
\why{apply} function that is generated during defunctionalization. A major
difference in our approach is that we do not represent continuations as an
algebraic data types. Instead, we fully capture of such values in their
specification, as we explain in what follows.

\subsection{Specifying continuations}
\label{sec:spec-cont}

Using all these different pieces, we can now conclude our \whyml proof by
creating a value \whyf{k} of type \whyf{continuation} and specifying it using
the \whyf{pre} and \whyf{post} predicates defined previously. For the example of
the \whyf{XCHG} handler, this is done as follows:
\begin{why3}
  try xchg (xchg 42) with
  | XCHG n ->
      val k : continuation int int in
      assume { ... };
      let old_p = !p in
      p := n;
      continue k old_p
\end{why3}
The \whyml \whyf{val..in} construction allows one to introduce local abstract
values, \emph{i.e.}, without imposing a particular implementation. The
conditions that go in the \whyf{assume} clause are essentially the specification
for~\whyf{k}. First, we state that the continuation is valid:
\begin{why3}
  assume { k._valid }
\end{why3}
Next, using the \whyf{pre} predicate, we state that the precondition for
\whyf{k} is the protocol's \emph{postcondition}. Although this might seem
counterintuitive at first, the protocol's postcondition is the condition that
\server must respect when it surrenders control to \client via the suspended
continuation. Therefore, the \client must ensure the protocol's postcondition
when calling the continuation. To recap, the protocol's postcondition is the
following:
\begin{gospel}
  (*@ protocol XCHG n :
      ensures reply = old !p && !p = n *)
\end{gospel}
We must now translate the postcondition to \whyml and encode it with the
\whyf{pre} predicate, as follows:
\begin{why3}
  assume { let eff_p = !p in
           forall reply state.
           pre k reply state <->
           reply = eff_p && state._p = n }
\end{why3}
Within this clause, we state that the precondition of \whyf{k} is that the reply
must be equal to \whyf{eff_p}, the value \whyf{p} takes when the continuation is
created (in other words, when the effect is performed). Next, using the
\whyf{state} type, we say that when \whyf{k} is called the value stored in
\whyf{p} is equal to \whyf{n}, the value passed by \client.

Finally, using the \whyf{post} predicate, we state that the postcondition of
\whyf{k} is the handler's postcondition.  This is because calling \whyf{k}
reinstalls the handler, any call to \whyf{k} terminates when it exists the
handler. Once again, the postcondition of the handler is
\begin{gospel}
  try_ensures !p = old !p && result = 42
\end{gospel}
To translate this to \whyml, we need to access the value \whyf{p} held before
\whyf{xchg} was called. To do so, we create an auxiliary variable \whyf{init_p}
and translate the handler's postcondition as follows:
\begin{why3}
  let init_p = !p in
  try xchg (xchg 42) with
  | XCHG n ->
    ...
    assume { forall reply old_state state result.
             post k reply old_state state result <->
             state._p = 3 && result = init_p }
\end{why3}
It is worth noting that we do not use the initial state in which the
continuation is called nor the reply its its postcondition. This is because
these values are only relevant for the continuation's precondition.

\paragraph{The generated Verification Conditions.} We take a step back and
observe what VCs \why generates for the resulting \whyml program, and what they
mean from the perspective of the original \ocaml program. For \client, \why
generates one VC checking if the postcondition of the \whyf{xchg} function
holds. Seeing as this function has the same postcondition as the
\whyf{perform_XCHG} function, this is proved trivially. As for \server, we
generate four VCs. First, one must prove that the handler's invariant holds when
the function terminates through the non-exceptional branch. Since the handler's
invariant is the same as \whyf{xchg}'s postcondition, this is also proved. Next,
the proof that the continuation has not been called yet (trivially, since the
continuation is created valid) and that the continuation's precondition is
respected. Given this handler respects the protocol, this VC is also
dispatched. Finally the proof that the handler's invariant is maintained when we
exit through the exceptional case. Seeing as the handler's last instruction is
calling the continuation, whose postcondition is the handler's invariant, this
is also proved.

We give in Appendix~\ref{sec:gener-transl-scheme} a general presentation of our
translation scheme from \ocaml programs with effects and handlers to \whyml. We
believe this approach is general enough, hence it could be adapted to any
proof-aware language with exceptions.

\section{Evaluation}
\label{sec:evaluation}

\begin{figure}[t]
  \centering
  \begin{tabular}
    {>{\raggedright}m{.33\textwidth}|>
    {\centering}m{.1\textwidth}|>
    {\centering}m{.275\textwidth}|c}
    Case study               & \# VCs & LOC / Spec. / Ghost & Proof time \\\hline
    Control inversion        & 15     & 20 / 25 / 2         & 0.74   \\
    \rowcolor{thegray}
    Division interpreter     & 19     & 18 / 16 / 2         & 0.74   \\
    \koda algorithm          & 142    & 26 / 20 / 0         & 17.60  \\
    \rowcolor{thegray}
    References               & 36     & 37 / 22 / 3         & 0.91 \\
    Xchg                     & 13     & 13 / 7  / 0         & 0.64 \\
    \rowcolor{thegray}
    \hline\hline
    Shallow Handlers         & 13     & 21 / 7  / 0         & 0.54 \\
  \end{tabular}
  \mycaption{Summary of the case studies verified with extended \cameleer.}
  \label{figure:case-studies}
\end{figure}

\mario{Verificar se queremos manter a coluna ``Immediate'', visto que o único
  exemplo que não é imediato é o do Koda Ruskey}

\mario{Dizer claramente que usamos nos case studies a sintaxe actual do OCaml}

\mario{Dar o link público para tudo.}

To test our approach, we specify and verify a handful of illustrative case
studies.  Besides the \whyf{XCHG} example we have already presented in detail,
we also verify the following case studies:
\begin{enumerate}
\item Turning a higher order iterator function into a generator. This process is
  commonly referred to as \emph{control inversion}.
\item A division interpreter that performs an effect in case of a division by zero.
\item Implementing mutable state using algebraic effects and deep handlers, as
  shown in Sec.~\ref{sec:algebraic-effects}.
\item The Koda-Ruskey algorithm.
\item Implementing a deep handler using a shallow handler. Our support for
  shallow handlers is still experimental, as we are currently a significant
  number of bureaucratic verification conditions to deal with unhandled effects.
\end{enumerate}
The reported case studies are all automatically verified, using a combination of
Alt-Ergo 2.4.2, CVC4 1.8, and Z3 4.8.6 SMT
solvers. Fig.~\ref{figure:case-studies} summarizes important metrics about our
verified case studies: the number of generated verification conditions for each
example; the total lines of \ocaml code, \gospellang specification, and lines of
ghost (these are also included in the number of \ocaml LOC), respectively; and
finally the time it takes (in seconds) to replay a proof. The runtimes were
measured by averaging over ten runs on a Lenovo Thinkpad X1 Carbon 8th
Generation, running Linux Mint 20.1, \ocaml 5.0.0, and \why 1.5.1.

All examples, including \ocaml implementation, \gospellang specification, and
\why proof sessions files are publicly
available\footnote{\url{https://github.com/mrjazzybread/cameleer/tree/effects/tests}}. It
is worth noting that throughout this paper we use a friendly syntax for effect
and handlers, which is not supported natively by the \ocaml compiler (this was
in fact the syntax proposed in an experimental-oriented switch of the
compiler). However, all of our case studies are written using the effects and
handlers library supported by \ocaml 5.0.0.

\begin{figure}
\begin{ocamlenv}
  effect Yield : unit

  let koda_ruskey (f : forest) =
    let rec enum_eff (f : forest) ((baton : forest list)[@ghost])
    = match f with
      | E -> perform Yield
      | N (i, l, r) ->
         if bits.(i) = White then begin
           enum_eff r baton;
           bits.(i) <- Black;
           try enum_eff l (r :: baton) with
           | Yield -> enum_eff r baton; continue k ()
         end else begin
           begin try enum_eff l (r :: baton) with
           | effect Yield -> enum_eff r baton; continue k () end;
           bits.(i) <- White;
           enum_eff r baton end
    (*@ requires valid_nums_forest f (length bits)
        requires disjoint_stack f baton
        requires any_forest f bits.elts
        ensures forall i.
                not mem_forest i f && not mem_stack i baton ->
                bits[i] = old bits[i]
        ensures inverse (Cons f baton) (old bits.elts) bits.elts
        variant f
        protocol Yield {
          requires valid_coloring f bits.elts
          requires forall i.
                   not mem_stack i baton -> bits[i] = old bits[i]
          requires unchanged baton (old bits.elts) bits.elts ||
                   inverse baton (old bits.elts) bits.elts
          ensures forall i.
                  not mem_stack i baton -> bits[i] = old bits[i]
          ensures inverse baton (old bits.elts) bits.elts
          modifies bits } *)
  in enum_eff f []
(*@ requires valid_nums_forest f (length bits)
    requires white_forest f bits.elts
    performs Yield *)
  \end{ocamlenv}
\caption{The Koda-Ruskey algorithm implementation and specification.}
\label{fig:koda-ruskey}
\end{figure}

\paragraph{A more elaborate case study.} Most of the proofs enumerated, although
not trivial, are relatively small scale examples. The outlier is the \koda
algorithm~\cite{KODA1993324} which enumerates the valid colorings of a forest of
n-ary trees. The complete \ocaml implementation and \gospellang for this
algorithm is depicted in Fig.~\ref{fig:koda-ruskey}. Very briefly put, the nodes
of a tree can either be colored White or Black. If a node is colored White, all
of its children must be White; if it is Black, there are no restrictions posed
on the color of its children. Each time the algorithm reaches a new coloring, an
effect is performed. The global array \whyf{bits} is used to represent each new
coloring.

Not only is this a very challenging proof, it is the only implementation we are
aware of for the \koda algorithm using algebraic effects. Although we are not
going to go over the entire specification, the key part is that the protocol for
this function is defined locally. This is because we want to state that,
whenever an effect is performed, the forest has a valid coloring. Since the
forest is passed as an argument, however, we must define the protocol within the
scope of the function to refer to its argument. Our proof, namely auxiliary
definitions and lemmas about forests and colorings, follows closely a previous
\why for a first-order, stack-based version of the Koda-Ruskey
algorithm~\cite{FilliatreP16}.

\section{Related Work}
\label{sec:related-work}

The most common type of static verifications programmers use for their programs
is type-checking.  A possible way of extending a type system's reach is by
adding an effect system~\cite{eff-sys}.  In normal type systems, we generally
have type signatures such as $f :\alpha \xrightarrow{} \beta$, where $f$ is a
function that receives an argument of type $\alpha$ and returns a value of type
$\beta$.  When using an effect system, type signatures have an additional
parameter: a row $\rho$.  A row is simply a list of signatures indicating which
effects this function performs.  With this typing information, we can determine,
at compile time, if a program will perform any unhandled effects, thereby
removing yet another common source of errors.  Effect systems can be used to
track not only algebraic effects, but any kind of impure behavior such as
mutable state and divergence~\cite{koka}. This means we can imbue within the
typing information whether or not a function is pure.

Although effect handlers are, still, relatively niche, there has been some work
in modelling their behavior using equational theory. The vanguard of this
approach was Plotkin and Power~\cite{plotkin-og}~where they reason over effects
as equations of effectful computations. Some research has been done by applying
this algebraic approach to effects and handlers. For example, Plotkin and
Pretnar~\cite{plotkin, plotkin-3-the-revenge} developed a logic where the
semantics of effects was captured using this approach and handlers were verified
by determining if they satisfied a set of equalities. Additionally, some
implementations have been developed that embed these reasoning rules into \coq,
an interactive theorem prover~\cite{freespec, intree}.

Another important line of research is reasoning about the concurrent behavior of
effects and handlers into Separation Logic. For instance, Timany and
Birkedal~\cite{callcc} developed a framework for the verification of full stack
continuations in a language with \emph{call/cc}. Additionally, Hinrichsen
\textit{et al.}~\cite{protocol-og} developed an extension of
\coq~\cite{jung2018iris} for reasoning over programs with message passing using
\emph{protocols} that define the rules that two agents must respect in their
exchange. Drawing inspiration from this research, de Vilhena and
Pottier~\cite{sep-eff} developed a simplified version of protocols for programs
with algebraic effects. Their framework interprets effects and effect handlers
as a communication between two agents: the one who raised the effect and its
handler. To model this communication, they use protocols which describe the
values that can be sent in either direction during this communication. The logic
that governs these protocols is built on top of \iris, a higher order separation
logic encoded in \coq. These protocols are, indeed, the main inspiration for the
\cameleer extension we have developed.

\section{Conclusions and Future Work}
\label{sec:concl-future-work}

We have presented an extension to the \cameleer tool to specify and reason about
algebraic effects and handlers. This lead us to introduce several extensions to
the \gospellang itself, namely the introduction of protocols at the level of
specification, as well as the \whyf{performs} and \whyf{try_ensures} clauses. In
order to explore such specification approach, we implemented inside \cameleer a
translation scheme from \ocaml into a \whyml embedding, which features
exceptions and defunctionalization of continuations as a means to reason about
effects and handlers. To the best of our knowledge, this extension to \cameleer
is the first framework that proposes an automated verification path to algebraic
effects.

As future work, we plan extend our tool to handle hidden state, seeing as we can
only deal with mutable state defined at the top-level. Additionally, our tool
can only prove handlers that remain installed when we call the generated
continuation: we plan to add support for handlers where the continuation may
produce an unhandled effect. Finally, our long term goal is to employ \cameleer
in the effort of verifying larger \ocaml libraries that make a significant use
of effects and handlers. A good candidate is the
\texttt{eio}\footnote{\url{https://github.com/ocaml-multicore/eio}}, a library
that proposes direct-style concurrency primitives (implemented as effects and
handlers) for the \ocaml language.

\bibliography{bibliography.bib}

\begin{thebibliography}{10}
\providecommand{\url}[1]{\texttt{#1}}
\providecommand{\urlprefix}{URL }
\providecommand{\doi}[1]{https://doi.org/#1}

\bibitem{eff-sys}
Bauer, A., Pretnar, M.: An effect system for algebraic effects and handlers.
  In: Heckel, R., Milius, S. (eds.) Algebra and Coalgebra in Computer Science.
  pp. 1--16. Springer Berlin Heidelberg, Berlin, Heidelberg (2013)

\bibitem{why}
Bobot, F., Filliâtre, J.C., Marché, C., Paskevich, A.: Why3: Shepherd your
  herd of provers. Boogie 2011: First International Workshop on Intermediate
  Verification Languages  (05 2012)

\bibitem{ChargueraudFLP19}
Chargu{\'{e}}raud, A., Filli{\^{a}}tre, J., Louren{\c{c}}o, C., Pereira, M.:
  {GOSPEL --- Providing OCaml with a Formal Specification Language}. In: Formal
  Methods - The Next 30 Years - Third World Congress. Lecture Notes in Computer
  Science, vol. 11800, pp. 484--501. Springer (2019),
  \url{10.1007/978-3-030-30942-8\_29}

\bibitem{DanvyN01}
Danvy, O., Nielsen, L.R.: Defunctionalization at work. In: Proceedings of the
  3rd international {ACM} {SIGPLAN} conference on Principles and practice of
  declarative programming, September 5-7, 2001, Florence, Italy. pp. 162--174.
  {ACM} (2001). \doi{10.1145/773184.773202},
  \url{https://doi.org/10.1145/773184.773202}

\bibitem{sep-eff}
{de Vilhena}, P.E., Pottier, F.: A separation logic for effect handlers. Proc.
  ACM Program. Lang.  \textbf{5}(POPL) (jan 2021). \doi{10.1145/3434314},
  \url{https://doi.org/10.1145/3434314}

\bibitem{FilliatreP16}
Filli{\^{a}}tre, J., Pereira, M.: Producing all ideals of a forest, formally
  (verification pearl). In: Blazy, S., Chechik, M. (eds.) Verified Software.
  Theories, Tools, and Experiments - 8th International Conference, {VSTTE}
  2016, Toronto, ON, Canada, July 17-18, 2016, Revised Selected Papers. Lecture
  Notes in Computer Science, vol.~9971, pp. 46--55 (2016).
  \doi{10.1007/978-3-319-48869-1\_4},
  \url{https://doi.org/10.1007/978-3-319-48869-1\_4}

\bibitem{shallow}
Hillerstr{\"o}m, D., Lindley, S.: Shallow effect handlers. In: Ryu, S. (ed.)
  Programming Languages and Systems. pp. 415--435. Springer International
  Publishing, Cham (2018)

\bibitem{protocol-og}
Hinrichsen, J.K., Bengtson, J., Krebbers, R.: Actris: Session-type based
  reasoning in separation logic. Proc. ACM Program. Lang.  \textbf{4}(POPL)
  (dec 2019). \doi{10.1145/3371074}, \url{https://doi.org/10.1145/3371074}

\bibitem{jung2018iris}
Jung, R., Krebbers, R., Jourdan, J.H., Bizjak, A., Birkedal, L., Dreyer, D.:
  {Iris From the Ground Up: A Modular Foundation For Higher-order Concurrent
  Separation Logic}. {Journal of Functional Programming}  \textbf{28}(e20)
  (2018). \doi{10.1017/S0956796818000151},
  \url{https://hal.archives-ouvertes.fr/hal-01945446}

\bibitem{action}
Kammar, O., Lindley, S., Oury, N.: Handlers in action. SIGPLAN Not.
  \textbf{48}(9),  145–158 (sep 2013). \doi{10.1145/2544174.2500590},
  \url{https://doi.org/10.1145/2544174.2500590}

\bibitem{KanigFilliatre09wml}
Kanig, J., Filli\^atre, J.C.: {Who: A Verifier for Effectful Higher-order
  Programs}. In: ACM SIGPLAN Workshop on ML. Edinburgh, Scotland, UK (Aug
  2009), \url{http://www.lri.fr/~filliatr/ftp/publis/wml09.pdf}

\bibitem{KODA1993324}
Koda, Y., Ruskey, F.: A gray code for the ideals of a forest poset. Journal of
  Algorithms  \textbf{15}(2),  324--340 (1993).
  \doi{https://doi.org/10.1006/jagm.1993.1044},
  \url{https://www.sciencedirect.com/science/article/pii/S0196677483710448}

\bibitem{koka}
Leijen, D.: Koka: Programming with row polymorphic effect types. Electronic
  Proceedings in Theoretical Computer Science  \textbf{153} (06 2014).
  \doi{10.4204/EPTCS.153.8}

\bibitem{freespec}
Letan, T., R{\'e}gis-Gianas, Y., Chifflier, P., Hiet, G.: {Modular Verification
  of Programs with Effects and Effect Handlers in Coq}. In: {FM 2018 - 22nd
  International Symposium on Formal Methods}. LNCS, vol. 10951, pp. 338--354.
  {Springer}, Oxford, United Kingdom (Jul 2018).
  \doi{10.1007/978-3-319-95582-7\_20}, \url{https://hal.inria.fr/hal-01799712}

\bibitem{pereiracav21}
Pereira, M., Ravara, A.: Cameleer: {A} deductive verification tool for ocaml.
  In: Silva, A., Leino, K.R.M. (eds.) Computer Aided Verification - 33rd
  International Conference, {CAV} 2021, Virtual Event, July 20-23, 2021,
  Proceedings, Part {II}. Lecture Notes in Computer Science, vol. 12760, pp.
  677--689. Springer (2021). \doi{10.1007/978-3-030-81688-9\_31},
  \url{https://doi.org/10.1007/978-3-030-81688-9\_31}

\bibitem{plot}
Plotkin, G., Power, J.: Algebraic operations and generic effects. Applied
  Categorical Structures  \textbf{11},  69--94 (02 2003).
  \doi{10.1023/A:1023064908962}

\bibitem{plotkin-og}
Plotkin, G., Power, J.: Computational effects and operations: An overview.
  Electronic Notes in Theoretical Computer Science  \textbf{73},  149--163
  (2004). \doi{https://doi.org/10.1016/j.entcs.2004.08.008},
  \url{https://www.sciencedirect.com/science/article/pii/S1571066104050893},
  proceedings of the Workshop on Domains VI

\bibitem{plotkin}
Plotkin, G., Pretnar, M.: A logic for algebraic effects. In: Proceedings of the
  2008 23rd Annual IEEE Symposium on Logic in Computer Science. p. 118–129.
  LICS '08, IEEE Computer Society, USA (2008). \doi{10.1109/LICS.2008.45},
  \url{https://doi.org/10.1109/LICS.2008.45}

\bibitem{plotkin-3-the-revenge}
Plotkin, G., Pretnar, M.: Handlers of algebraic effects. In: Castagna, G. (ed.)
  Programming Languages and Systems. pp. 80--94. Springer Berlin Heidelberg,
  Berlin, Heidelberg (2009)

\bibitem{Regis-GianasP08}
R{\'{e}}gis{-}Gianas, Y., Pottier, F.: A hoare logic for call-by-value
  functional programs. In: Audebaud, P., Paulin{-}Mohring, C. (eds.)
  Mathematics of Program Construction, 9th International Conference, {MPC}
  2008, Marseille, France, July 15-18, 2008. Proceedings. Lecture Notes in
  Computer Science, vol.~5133, pp. 305--335. Springer (2008).
  \doi{10.1007/978-3-540-70594-9\_17},
  \url{https://doi.org/10.1007/978-3-540-70594-9\_17}

\bibitem{defunc}
Reynolds, J.C.: Definitional interpreters for higher-order programming
  languages. In: Proceedings of the ACM Annual Conference - Volume 2. p.
  717–740. ACM '72, Association for Computing Machinery, New York, NY, USA
  (1972). \doi{10.1145/800194.805852},
  \url{https://doi.org/10.1145/800194.805852}

\bibitem{eff-t}
Sivaramakrishnan, K., Dolan, S., White, L., Kelly, T., Jaffer, S.,
  Madhavapeddy, A.: Retrofitting effect handlers onto ocaml. In: Proceedings of
  the 42nd ACM SIGPLAN International Conference on Programming Language Design
  and Implementation. p. 206–221. PLDI 2021, Association for Computing
  Machinery, New York, NY, USA (2021). \doi{10.1145/3453483.3454039},
  \url{https://doi.org/10.1145/3453483.3454039}

\bibitem{callcc}
Timany, A., Birkedal, L.: Mechanized relational verification of concurrent
  programs with continuations. Proc. ACM Program. Lang.  \textbf{3}(ICFP) (jul
  2019). \doi{10.1145/3341709}, \url{https://doi.org/10.1145/3341709}

\bibitem{intree}
Xia, L.y., Zakowski, Y., He, P., Hur, C.K., Malecha, G., Pierce, B.C.,
  Zdancewic, S.: Interaction trees: Representing recursive and impure programs
  in coq. Proc. ACM Program. Lang.  \textbf{4}(POPL) (dec 2019).
  \doi{10.1145/3371119}, \url{https://doi.org/10.1145/3371119}

\end{thebibliography}
\bibliographystyle{splncs04}

\newpage

\appendix

\section{General Translation Scheme}
\label{sec:gener-transl-scheme}

\begin{figure}
$\h(e_{0t}$, $\overline{E\xbar}$, $\bar{e}$, $\overline{{S_h}_{ensures}}$,$term_{post}$, $\tau$, $\Upsigma$, $term_{state}$) =
\begin{why3}
  let handler () =
    ensures{${S_h}_{ensures_1}$}
    ensures{${S_h}_{ensures_2}$}
    ...
        let init_state = $\svars$ in
        try $e_{0t}$ with
        |...
        |$E_n\bar{x}$ ->
            let eff_state = $\svars$ in
            val gen_k unit : continuation $(\pi_2(\Upsigma(E_n)))$ $\tau$
            ensures{valid result}
            ensures{
                forall arg state.
                pre result arg state <->
                E_post arg eff_state state result
            }
            ensures{
                let f = result in
                forall arg irrelevant_old_state state result.
                post f arg irrelevant_old_state state result <->
                  let state_old = init_state in
                  $\mathit{term_{post} \wedge term_{state}}$
            } in let k = gen_k () in $e_{nt}$
         |...
    in handler ()
\end{why3}
\caption{Translation rule for effect handlers.}
\label{fig:try-trans}
\end{figure}

\begin{figure}
  \begin{align*}
    \iamgoingtokillyou(\epsilon) &= \mathtt{true} \\
    \iamgoingtokillyou(x::t) &= \mathtt{state}.x = \mathtt{state\_old}.x \wedge \iamgoingtokillyou(t)
  \end{align*}
\caption{Unmodified State}
\end{figure}

\begin{figure}

  \begin{align*}
    \mathcal{T}\ &(\tau_1 \xrightarrow{} \tau_2 \xrightarrow{} \tau_3)=\\
    & let\ s = \mathcal{T}(\tau_2\xrightarrow{} \tau_3)\ in\\
    & \tau_1::\pi_1(s), \pi_2(s)\\\\
    \mathcal{T}\ &(\tau_1 \xrightarrow{} \tau_2 \xrightarrow{} \tau_3)=\\
    \ &(\tau_1 \xrightarrow{} \tau_2) = \tau_1, \tau_2 \\ \\
    \mathcal{T}\ &(\tau_1 \xrightarrow{} \tau_2 \xrightarrow{} \tau_3)=\\
    \mathcal{T}\ &(\tau) = unit, \tau
  \end{align*}
  \caption{Effect Type Translation.}
  \label{fig:eff-trans}
  \end{figure}

\begin{figure}
  $\df\ (args = x_1$ : $\tau_1$, $x_2$ : $\tau_2$, ..., $x_n$ : $\tau_n$) $\tau_r$ $e_t$ $\bar{S}$\ $term_{pre}$\ $term_{post}$ =
  \begin{why3}
      let f $args$ $\bar{S}$ = $e_t$ in
      val gen_f () : lambda $(\tau_1, \tau_2, ..., \tau_n)$ $\tau_r$
      ensures{
        forall arg state. pre result arg state <->
          let $x_1$, $x_2$ ... = arg in
          $term_{pre}$
      }
      ensures{
        let f = result in
        forall arg old_state state result.
        post f arg old_state state result <->
        let $x_1$, $x_2$ ... = arg in
        $term_{post}$
      } in gen_f ()
  \end{why3}

\caption{Defunctionalization Function.}
\label{fig:def-rule}
\end{figure}

\begin{figure}
  \begin{align*}
    \s(term::xs)  &= h(term) \wedge \s(xs)\\
    \s(\epsilon) &= true
  \end{align*}
\caption{Combination of terms}
\end{figure}

\begin{figure}
  $\mathcal{P}(E,\ \xbar,\ \Upsigma,\ term_{pre},\ term_{post},\ mod) =$
  \begin{why3}
      predicate pre_$E$ (arg : $\env{1}$) (state : state) =
          let $\xbar$ = arg in
          $term_{pre}$

      predicate post_$E$ (arg : $\env{1}$) (old_state : state)
                        (state : $\stype$) (reply : $\env{2}$) =
          let $\xbar$ = arg in
          $term_{post}$


      val perform_$E$ (arg: $\env{1}$) : $\env{2}$
      requires{pre_E arg $\svars$}
      ensures{post_E arg (old $\svars$) $\svars$ result}
      writes{$mod$}
  \end{why3}
  \caption[]{Protocol Translation Function.}
  \label{fig:trans-rules}
  \end{figure}

\begin{figure}
  \small
  \centering
  \begin{judge}[b]{\textwidth}
    \hfill

\[
  \begin{array}{c}
  \inferrule*[Right=\teffect]
  { }
  {\trans
      {\mathtt{effect}\ E\ :\ \tau}
      {\mathtt{exception}\ E}
      [\Upsigma[E\xrightarrow{}\mathcal{T}(\tau)]]
  }\\\\

    \inferrule*[Right=\tperform]
    { }
    {\trans
        {\mathtt{perform\ E\xbar}}
        {\mathtt{perform\_E\ \xbar}}
    }\\\\

    \inferrule*[Right=\tprotocol]
    {\s(\overline{Sp}_{requires}) = term_{pre}\\
     \s(\overline{Sp}_{ensures}) = term_{post}
    }
    {\trans
        {\mathtt{protocol}\ E\xbar : \overline{Sp}}
        {\mathcal{P}(E,\ \xbar,\ \Upsigma,\ term_{pre},\ term_{post},\ Sp_{modifies})}
    } \\\\

    \inferrule*[Right=\tfun]
    {\trans[\mkenv{\effenv}{\funenv}{\argenv\cup\overline{x}}{\emptyset}]{e}{e_t}[\mkenv{}{}{}{\refenv'}]\\ \s(\bar{S}_{requires})=pre \\ \s({\bar{S}}_{ensures}) = post}
    {\trans{\mathtt{fun}\ \bar{S}\ \overline{(x : \tau_{arg})} : \tau_{ret} \xrightarrow{} e}
    {\df\ \overline{(x: \tau_{arg})}\ \tau_{ret}\ e_t\ \bar{S}\ pre\ post}}\\\\

    \inferrule*[Right=\tperformsclause]
    { }
    {\trans
        {\mathtt{performs}\ E}
        {\mathtt{raises}\{E\ arg\xrightarrow{} E\_pre\ arg\ \svars\}}
    }\\\\

    \inferrule*[Right=\tappdefun]
    {\forall i\ \trans[\mkenv{s}{s}{s}{\refenv_i}]{e_i}{e_{it}}[\mkenv{}{}{}{\refenv_{i+1}}] \\ e_0 = x \implies x\in\argenv}
    {\trans[\mkenv{s}{s}{s}{\refenv_0}]{e_0\ e_1\ ...\ e_n}{\mathtt{apply}\ e_0\ (e_2, e_3, ..., e_n)}[\mkenv{}{}{}{\allvars}]}\\\\

  \inferrule*[Right=\tapp]
    {\forall i\ \trans[\mkenv{s}{s}{s}{\refenv_i}]{e_i}{e_{it}}[\mkenv{}{}{}{\refenv_{i+1}}] \\ \neg x\in\argenv}
    {\trans{x\ e_0\ e_1\ ...\ e_n}{x\ e_{0t}\ e_{1t}\ \dots e_{nt}}[\funenv(x)\cup\mkenv{}{}{}{\refenv_{n + 1}}]} \\\\

    \inferrule*[Right=(TTry)]
    {\forall i. 0 < i \leqq n \implies \ \trans[\mkenv{s}{s}{s}{\refenv_i}]{e_i}{e_{it}}[\mkenv{}{}{}{\refenv_{i+1}}] \\ \s(\overline{Sh_{ensures}}) = term_{post} \\ Sh_{returns} = \tau \\ e_{try} = try\ e_0\ with\ Sh \ \overline{\mathtt{effect}\ E\bar{x}\ k\xrightarrow{}e}}
    {\trans
      [\mkenv{s}{s}{s}{\emptyset}]
      {e_{try}}
      {\h(\overline{E\bar{x}},\ \overline{e},\ \overline{Sh_{ensures}},\ term_{post},\ \tau,\ \Upsigma,\ \iamgoingtokillyou(\allvars\setminus\refenv_{n+1}))}[\mkenv{}{}{}{\refenv_{n}}]} \\\\

    \inferrule*[Right=(TLet)]
    {\trans[\mkenv{s}{s}{s}{\emptyset}]{e}{e_t}[\mkenv{}{}{}{\refenv}] \\
     e\neq fun\ ... \\ \bar{x} \neq \epsilon \implies \funenv' = \funenv[f\xrightarrow{}\refenv] \\
     \bar{x} = \epsilon \implies \funenv' = \funenv}
    {\trans[\mkenv{s}{s}{s}{\emptyset}]
        {\mathtt{let}\ \bar{S}\ f\ \bar{x} = e}
        {\mathtt{let}\ \bar{S}\ f\ \bar{x} = e_t}[\mkenv{}{\funenv'}{}{}]
    }\\\\

    \inferrule*[Right=(TIf)]
        {\trans{e_1}{e_{1t}}[\mkenv{}{}{}{\refenv'}] \\ \trans[\mkenv{s}{s}{s}{\refenv'}]{e_2}{e_{2t}}[\mkenv{}{}{}{\refenv''}] }
        {\trans
            {\mathtt{if}\ a\ \mathtt{then}\ e_{1}\ \mathtt{else}\ e_{2}}
            {\mathtt{if}\ a\ \mathtt{then}\ e_{1t}\ \mathtt{else}\ e_{2t}}[\mkenv{}{}{}{\refenv''}]
        }

    \\\\
    \inferrule*[Right=(TSeq)]
        {\trans{e_1}{e_{1t}}[\mkenv{}{}{}{\refenv'}] \\ \trans[\mkenv{s}{s}{s}{\refenv'}]{e_2}{e_{2t}}[\mkenv{}{}{}{\refenv''}]}
        {\trans{e_1\ ;\ e_2}{e_1\ ;\ e_2}[\mkenv{}{}{}{\refenv''}]}
    \\\\

    \inferrule*[Right=(TLetIn)]
    {\bar{x} \neq \epsilon \implies \refenv = \emptyset \\ \trans[\mkenv{s}{s}{s}{\refenv}]{e_1}{e_{1t}}[\mkenv{}{}{}{\refenv'}] \\
     e\neq fun\ ... \\ \bar{x} \neq \epsilon \implies \funenv' = \funenv[f\xrightarrow{}\refenv'] \\
     \bar{x} = \epsilon \implies \funenv' = \funenv \\ \trans[\mkenv{s}{\funenv'}{s}{\refenv'}]{e_2}{e_{2t}}[\mkenv{}{}{}{\refenv''}]}
    {\trans
        {\mathtt{let}\ \bar{S}\ f\ \bar{x} = e_1\ \mathtt{in}\ e_2}
        {\mathtt{let}\ \bar{S}\ f\ \bar{x} = e_{1t}\ \mathtt{in}\ e_{2t} }[\mkenv{}{}{}{\refenv''}]
    }\\\\

    \inferrule*[Right=(TMatch)]
        {\forall i. 0 \leqq i < n \implies \trans[\mkenv{s}{\funenv}{s}{\refenv_i}]{e_i}{e_{it}}[\mkenv{s}{s}{s}{\refenv_{i+1}}]}
        {\trans{\mathtt{match}\ a\ \mathtt{with}\ \overline{p\xrightarrow{}e_i}}{\mathtt{match}\ a\ \mathtt{with}\  \overline{p\xrightarrow{}e_{it}}}[\mkenv{s}{s}{s}{\refenv_{n}}]}
    \\\\

    \inferrule*[Right=(TModifies)]
        { }
        {\trans{\mathtt{modifies}\ \bar{s}}{\mathtt{writes}\{s\}}}\qqquad\qquad

    \inferrule*[Right=(Tempty)]
        {}
        {\trans{\epsilon}{\epsilon}}\\\\

    \inferrule*[Right=(TDecl)]
        {\trans[\mkenv{s}{s}{s}{\emptyset}]{d}{d_t}[\mkenv{\effenv'}{\funenv'}{}{}]\\\trans[\mkenv{\effenv'}{\funenv'}{s}{\emptyset}]{\bar{d}}{d_{tl}}[\mkenv{\effenv''}{\funenv''}{}{}]}
        {\trans[\mkenv{s}{s}{s}{\emptyset}]{d::\bar{d}}{d_t::d_{tl}}[\mkenv{\effenv''}{\funenv''}{}{}]}
  \end{array}
\]
  \end{judge}
  \caption{Inductive Translation Rules.}
  \label{fig:scheme}
\end{figure}

\end{document}